# No signature of the orbital motion of a putative 70 solar mass black hole in LB-1

**Authors:** Michael Abdul-Masih[1], Gareth Banyard[1], Julia Bodensteiner[1], Dominic M. Bowman[1], Karan Dsilva[1], Matthias Fabry[1], Calum Hawcroft[1], Laurent Mahy[1], Pablo Marchant[1], Gert Raskin[1], Maddalena Reggiani[1], Hugues Sana[1], Tomer Shenar[1], Andrew Tkachenko[1], Hans Van Winckel[1], Lore Vermeylen[2]

**Affiliations:**
[1]Institute of Astrophysics, KU Leuven, Celestijnenlaan 200D, Leuven, Belgium
[2]Royal Observatory of Belgium, Ringlaan 3, 1180 Brussels, Belgium

**Correspondence** should be addressed to H.S. (hugues.sana@kuleuven.be)

**Abstract:** Liu et al.[1] recently reported the detection of a 68 [+11/-13] solar mass ($M_\odot$) black hole (BH) paired with an 8.2 [+0.9/-1.2] $M_\odot$ B-type sub-giant star in the 78.9-day spectroscopic binary system LB-1. Such a black hole is over twice as massive as any other known stellar-mass black hole with non-compact companions[2,3] and its mass approaches those that result from BH-BH coalescences that are detected by gravitational wave interferometers[4]. Its presence in a solar-like metallicity environment challenges conventional theories of massive binary evolution, stellar winds and core-collapse supernovae, so that more exotic scenarios seem to be needed to explain the existence and properties of LB-1[5,6]. Here, we show that the observational diagnostics used to derive the BH mass results from the orbital motion of the B-type star, not that of the BH. As a consequence, no evidence for a massive BH remains in the data, therefore solving the existing tension with formation models of such a massive BH at solar metallicity and with theories of massive star evolution in general.

The conclusion of a large BH mass relies on two main arguments:
(i) the characterization of the orbital and physical properties of the B-star companion;
(ii) an indirect measurement of the reflex orbital motion of the BH.

The critical measure that yields the high-mass for the hidden companion comes from the semi-amplitude $K_{BH}$ = 6.4 km s$^{-1}$ of the RV curve of the potential BH companion, a result that is measured from the apparent wobbling of the position of the wings of the H$\alpha$ emission profile, assuming this emission is generated by a passive accretion disc around this unseen companion. While Liu et al. assumed that such wobbling was tracing the BH reflex motion, here we show that the observed RV measurements results from the superposition of the B-star stellar absorption line superposed on a static H$\alpha$ emission profile. This can be demonstrated either observationally or with pure simulations.

To reach this conclusion, we used new high-spectral resolution observations of LB-1 obtained with the HERMES spectrograph[7] coupled with the Mercator telescope at La Palma, Spain (see Appendix). Using these new data, we isolate the pure H$\alpha$ emission profile by subtracting, from the observed profile, a theoretical H$\alpha$ absorption profile corresponding to the best-fit atmospheric parameters of the detected B-type star (see Appendix), after accounting for the orbital shift of the observed profile given the epoch of the new HERMES observations. The pure emission profile is displayed in Fig. 1. We then create a series of reconstructed H$\alpha$ profiles that combine the (static) emission profile and the phase-shifted H$\alpha$ absorption from the B-star companion at phases corresponding to the ones observed by Liu et al.[1] (Fig. 2).

Finally, we used a barycentre method as well as a bissectrice method with an identical mask as applied in Liu et al.[1], to estimate the apparent RV-shift resulting from the combined H$\alpha$ profile, obtaining similar results from both methods. The RV measurements are displayed in Fig. 2 and reveal a clear low-amplitude sinusoidal variation in perfect anti-phase with that of the B-type star. In addition, the derived (apparent) semi-amplitude is 5.4 km s$^{-1}$, in good agreement with that measured by Liu et al. (6.4 +/- 0.8 km s$^{-1}$) given the high spectral resolution of our experiment. Very similar results were obtained when repeating the procedure with a Gaussian profile representing the H$\alpha$ line instead of the one constructed from observations (see Appendix).

This demonstrates that the largest contribution of the detected RV signal obtained from the barycenter method applied to the H$\alpha$ profile is readily explained by the contamination of the strong H$\alpha$ emission component with the B-type star absorption component and is thus not able to trace the reflex orbital motion of the BH. Consequently, there is no evidence for a large mass-ratio and hence a large absolute mass of the BH. The remaining observational constraints are limited to the minimum mass of the companion to the B-type star, of the order of 4 M$_\odot$ after reassessing the physical parameters of the B-star companion (see Appendix) and a static or very low amplitude (<1km s$^{-1}$) emission in H$\alpha$.

Our results solve the apparent challenge posed by the presence of a very massive black hole at solar metallicity, in a mass range where pair-instability supernovae are expected with a very different end-product, as well as the absence of X-ray emission. Further simulations of composite spectra formed by two ~4 solar mass object main-sequence stars, where one of them is a rapidly rotating B-type stars shows that the latter, if rotating at projected rotational velocity larger than about 200 km s$^{-1}$, will be very difficult or impossible to detect given the quality of the data collected so far (see Appendix). Alternatively, it is still possible that a black hole is present in LB-1. There are however no observational data specifically requiring a very large (>50 M$_\odot$) mass.

Further avenues to obtain a self-consistent picture of the intriguing LB-1 binary system may lay in a static H$\alpha$ emission that is coming from a gaseous circumbinary disk similar to those seen in some B[e] stars[8] or a more massive counterpart of post-Asymptotic-Giant Branch phenomena[9]. Alternatively, it could be the progenitor of a low- to intermediate-mass X-ray binary, i.e. a lower-mass B-type star in a binary system with a black hole.

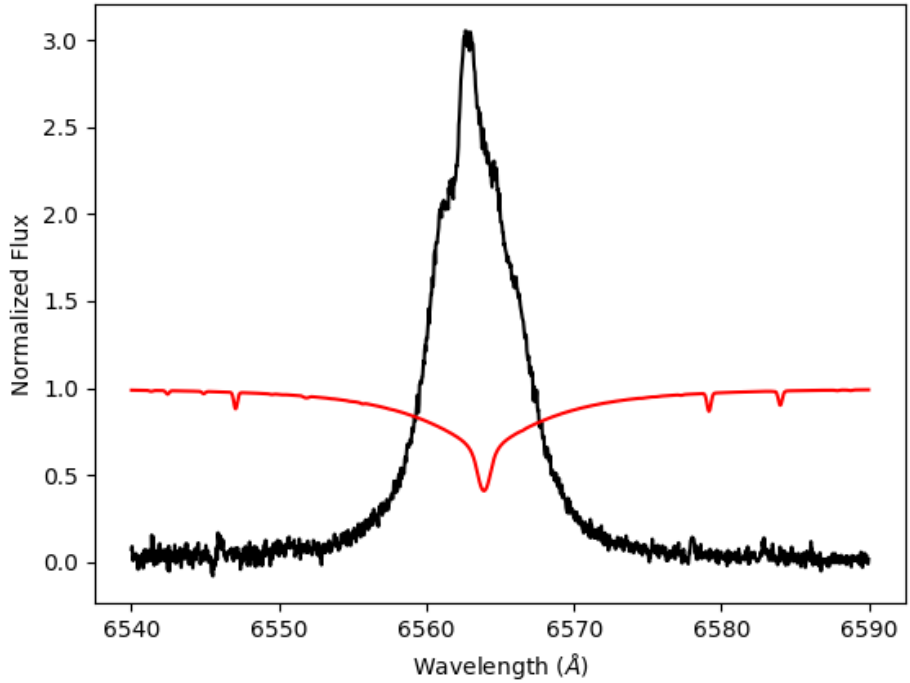

**Fig. 1.** Reconstructed Hα emission profile (black) obtained by subtracting the B-star stellar absorption component (red) corresponding to the best-fit atmospheric model (see Appendix) to the observed HERMES spectrum.

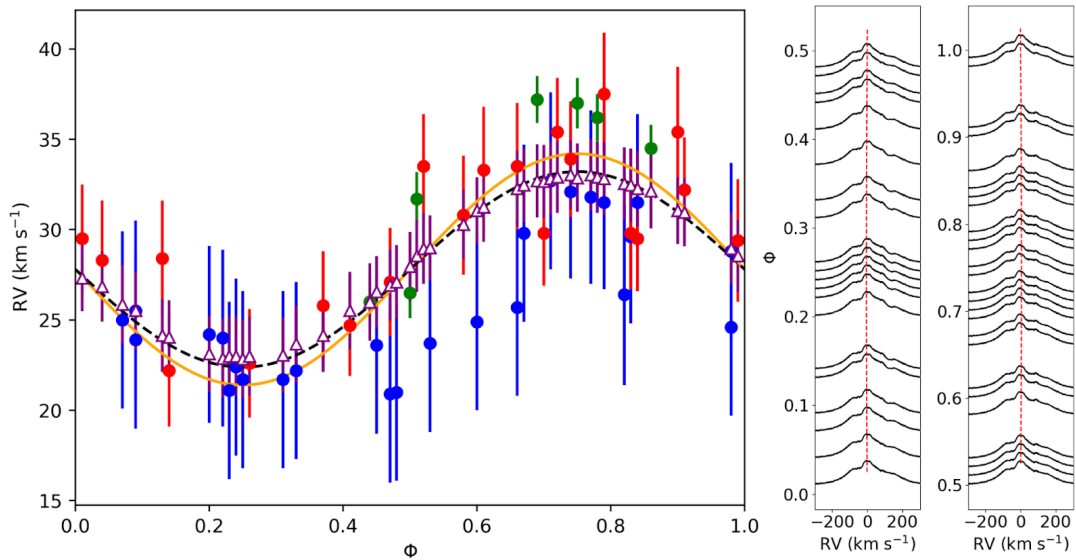

**Fig. 2. Left:** Filled symbols: Measured RVs from the barycenter method applied to the observed emission Hα profile (data and ephemeris adopted from Liu et al.[1]: Blue: LAMOST. Red: GTC. Green: Keck). Open symbols (purple): Results from our simulated profiles assuming a fixed emission profile and an over imposed Hα absorption line from the B-star component. The best fitted RV curves from the two datasets have semi-amplitudes of, respectively, 6.4 km s$^{-1}$ (plain line, orbital solution from [1]) and 5.4 km s$^{-1}$ (dashed lines, resulting from a fit to the open symbols), and are almost indistinguishable given the measurement errors. **Right:** Simulated Hα profiles formed by a static emission component and a Doppler-shifted stellar absorption according to the orbital phase ϕ and ephemeris from Liu et al.[1] (see individual components in Fig. 1).

# APPENDIX

## High-resolution spectroscopic observations

We acquired high-spectral resolution optical observations of LB-1 on 04.12.2019 using the HERMES spectrograph[7] at the Mercator telescope located at the Roque de Los Muchachos Observatory in La Palma, Spain. HERMES delivered the full optical range from 3800 to 9000 Angstroms at the spectral resolving power of 80,000. The observation sequence consisted of a pair of 1800 s exposures taken back to back. The data were reduced with the HERMES pipeline, co-added and normalized so that the continuum is at unity. The reduced and co-added spectrum yielded a signal-to-noise ratio (S/N) of 60. This is equivalent to an S/N of 70 when degrading to the resolving power of HIRES/Keck-I and a S/N of 400 when degrading to the resolving power of LAMOST. We further retrieved the HIRES data from the Keck archive to inform our analysis.

## Atmosphere Analysis

We determine stellar parameters using the Grid Search in Stellar Parameters fitting suite (GSSP)[10]. GSSP simultaneously fits the effective temperature ($T_{eff}$), surface gravity (log $g$), projected rotational velocity ($v_{eq} \sin i$), micro- and macro-turbulent velocities ($v_{micro}$ and $v_{macro}$, resp.) as well as metallicity ($Z$). A $\chi^2$ between the observed spectrum and the models is computed for a set of synthetic spectra calculated using the local-thermal-equilibrium (LTE) radiative transfer code SYNTHV[11] from a grid of LLmodel atmospheres[12].

When fitting the observed HERMES spectrum, we exclude wavelength regions that are definitely or possibly affected by emission. This includes H$\alpha$ but contamination by emission is also visible in H$\beta$, hence we exclude the cores of all Balmer lines. The results of the fit are not sensitive to the clipping of the lines but are sensitive to the way normalization is performed.

Our best-fit model (see Figs. A1 and A2) corresponds to an effective temperature $T_{eff}$ = 13,500 +/- 700 K, a surface gravity log $g$ = 3.3 +/- 0.3 [cgs] and a projected rotational velocity of $v_{eq} \sin i$ = 7.5 +/- 4.0 km s$^{-1}$. The metallicity is not well constrained to $Z$=0.6 [+0.4/-0.6] $Z_\odot$. Both micro- and macro-turbulent velocities are low, i.e. $v_{micro}$ = 0.2 km s$^{-1}$ and $v_{macro}$ = 4 km s$^{-1}$. Different normalizations yield slightly different results but do not change the estimated mass of the B-type star (see below).

While most of these values are in agreement with the results of Liu et al., we note a significantly lower effective temperature by almost 5000 K. We performed a similar fit using the TLUSTY BSTAR2006 grid[13] as adopted by Liu et al.[1] and we found that values of the order of 15,000 K are privileged. This, however, corresponds to the lower $T_{eff}$ boundary of the BSTAR2006 grid. It is therefore possible that this prevented Liu et al.[1] to converge to temperatures below 15,000 K. Alternatively, apparent difficulties in the normalization of the HIRES/KECK spectra may have led to different quantitative results. The implications of significantly different atmospheric parameters are important as, using the BONNSAI[14] Bayesian tool to compare our measurements with stellar evolutionary models at solar metallicity[15], we obtain a best-fit evolutionary mass for the B-type companion of 4.2 [+0.8,-0.7] M$_\odot$.

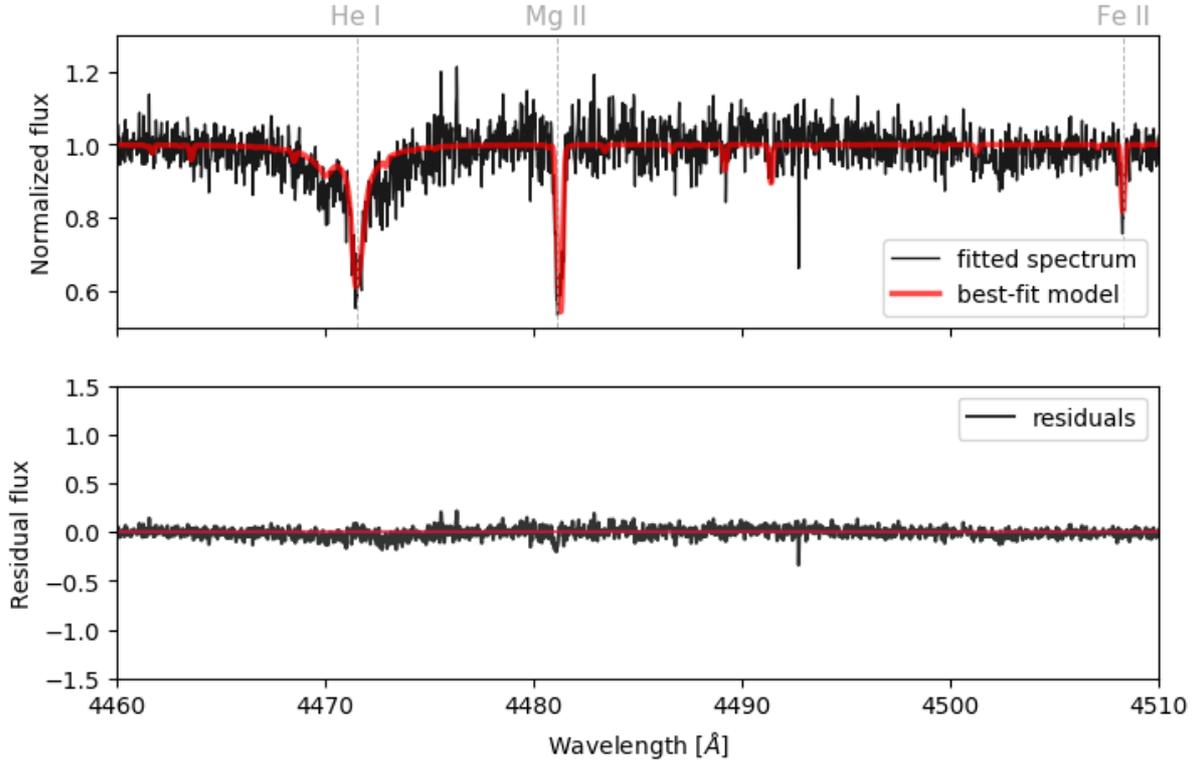

**Fig. A1.** Best GSSP atmosphere model overlaid on the observed HERMES spectrum (top) and residuals of the fit (bottom). Only a small portion of the spectrum is shown.

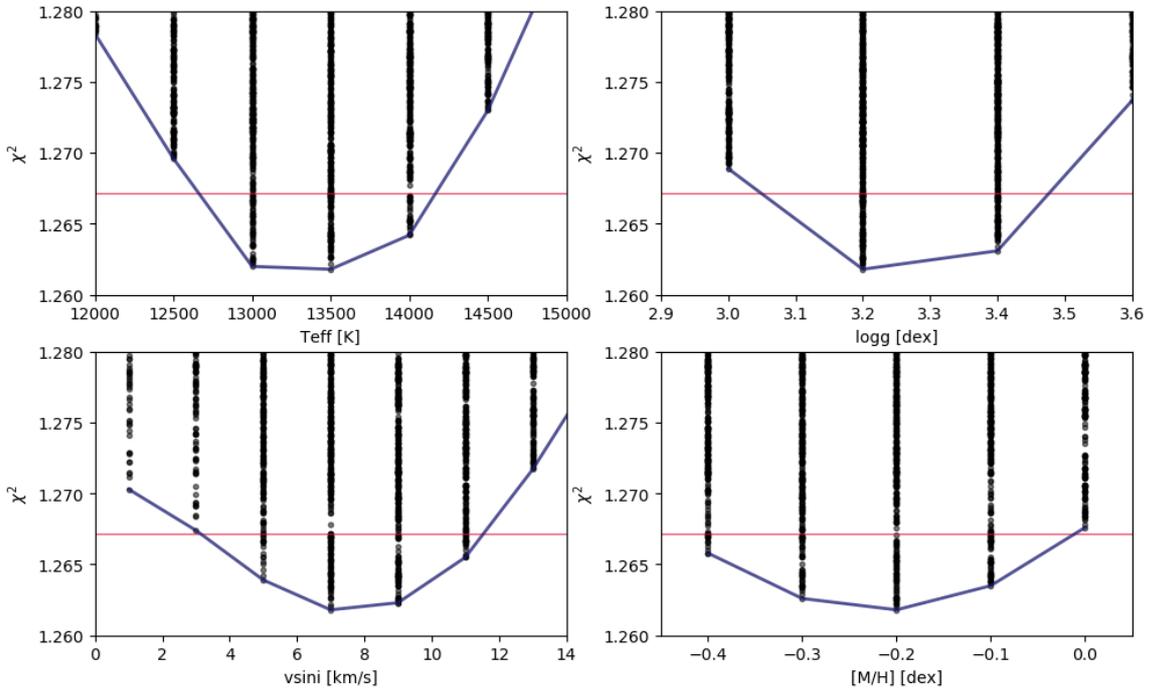

**Fig. A2.** Chi-squared distributions of the GSSP atmosphere fitting for effective temperature, surface gravity, projected rotational velocity, and metallicity from top left to bottom right, respectively. The black points are models, the blue line is an interpolation among the grid points, and the red line corresponds to the 1σ confidence level.

**Additional simulations on the impact of composite profiles on RV measurement**

The main approach used the observational data and is described in the text. Here we present additional simulations that reveal the same results without relying on any observations. Specifically, and to ensure that our method of constructing the Hα emission component through the combination of models and observations does not indirectly introduce the measured anti-phased trend, we repeat this experiment using synthetic models. For this purpose, the Hα line is modeled as a Gaussian profile with a standard deviation of 150 km s$^{-1}$ and a maximum amplitude of 3.5, i.e. similar to the observed properties. For the B-star model, we used a TLUSTY model with $T_{eff}$ = 15,000 K and log $g$ = 3.5 [cgs]. The models are degraded to a resolution of R=80,000 and a S/N=50.

We shifted the B-star model according to the orbital velocities derived by Liu et al.[1]. The Hα mask is maintained static. We then measured the RVs of the Hα line using precisely the method implemented by Liu et al.[1]. The simulated spectra at conjunction and quadratures are shown in Fig. A3.

In Fig. A4, we show the measured RVs of the Hα line compared with the B-star RVs, which are based on orbital solution derived by Liu et al.[1]. An anti-phased behavior similar to that detected by Liu et al.[1] is immediately apparent despite the fact that the emission profile is static in our experiment. We obtain a best-fit for a secondary amplitude of $K_2$ = 4.2 +/- 0.2 km s$^{-1}$. However, as Liu et al.[1] report, the amplitude strongly depends on the region in which Hα is considered. The results of this test are further evidence that the reported anti-phased behavior[1] is spurious to the RV method.

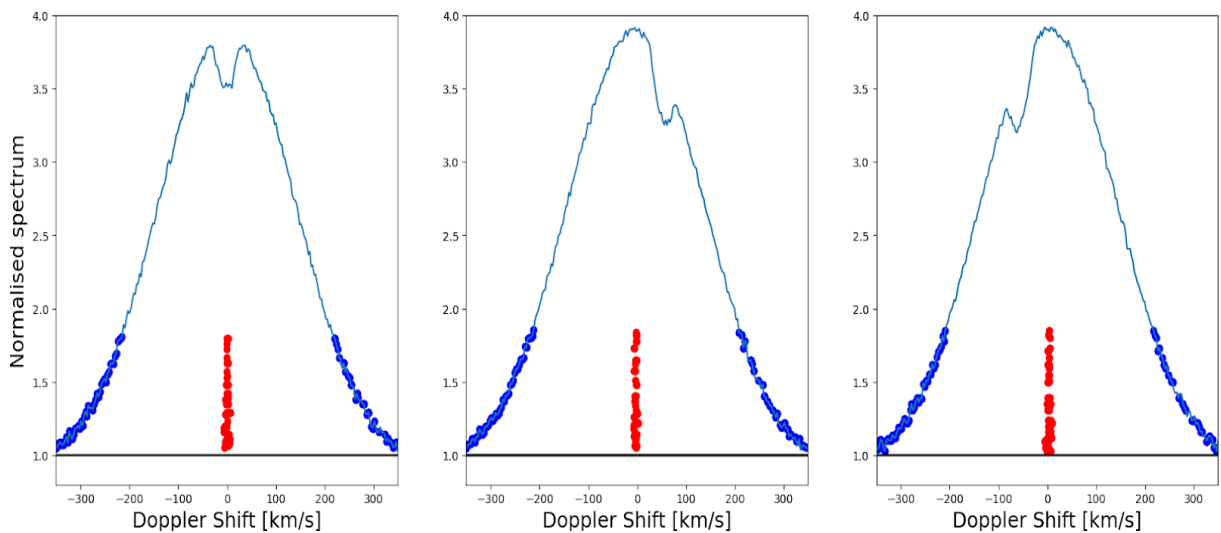

**Fig. A3:** Simulated noisy spectra (S/N = 50, R = 80,000) of Hα emission (Gaussian profile) co-added with a TLUSTY B-star spectrum at conjunction (left panel) and quadratures (middle and right panel). Marked are the bisectrice of the fitted region (⅓ of the profile, as in Liu et al. 2019)

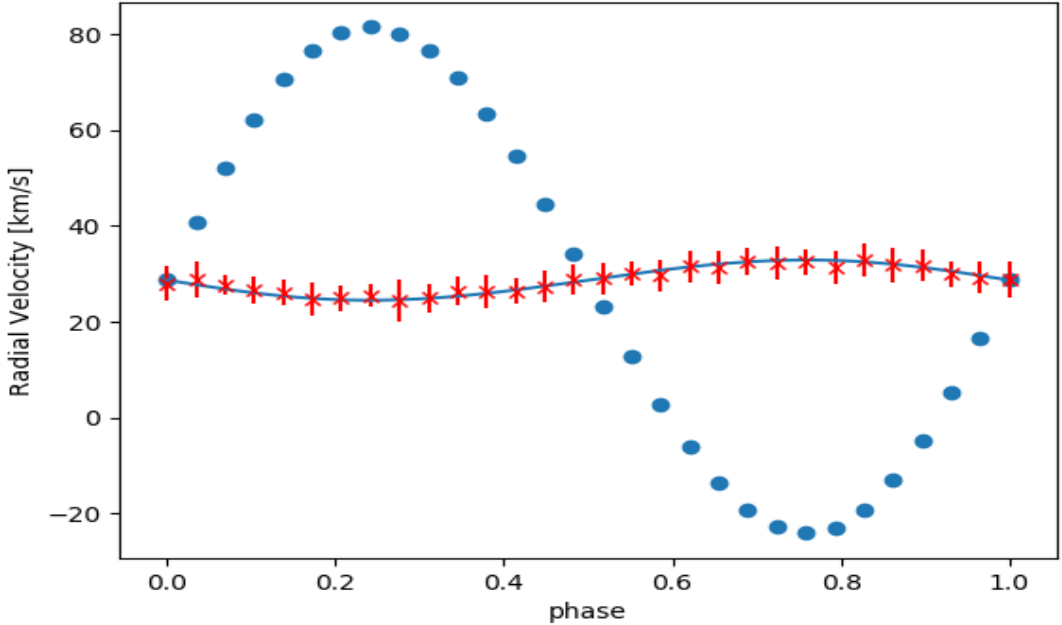

**Fig. A4.** RVs of the visible B-star companion computed from [1] (blue circles) compared with Hα RVs measured using the barycentre method (red crosses). A RV curve with $K_2$ = 4.2 km s$^{-1}$ is overlaid.

**Constraints on the nature of the companion**

Our analysis demonstrates that the value of $K_2$=6.4 km s$^{-1}$ reported by Liu et al. (2019) is a spurious result that does not constrain the orbital motion of the hidden companion. Using the known orbital parameters and the estimated mass of the visible B-type star (~4.2 M$_\odot$), one easily computes that the minimum mass of the companion is about 4 solar masses. We can therefore rule out a neutron star companion. Viable options are thus a rapidly-rotating main sequence star of a comparable mass to the B-type primary, a He star, or a black hole.

To consider the possibility of a rapidly rotating star having been missed in the available data, we simulated synthetic composite spectra using two TLUSTY models of identical parameters ($T_{eff}$ = 15,000K, log $g$ = 3.5 [cgs]) and an equal light ratio. We fix the model of the visible B-star to the observed broadening parameters ($v_{eq}$ sin $i$ = 7.5 km s$^{-1}$, $v_{macro}$ = 4 km s$^{-1}$). For the second model, we used various rotational broadening of $v_{eq}$ sin $i$ = 100, 200, and 300 km s$^{-1}$. The results are displayed in Fig. A5. We conclude that the presence of a main sequence star with $v_{eq}$ sin $i$ values above 200 km s$^{-1}$ is difficult to test without high S/N data (>200, judging by the depth of the lines) and careful normalisation.

It is interesting to point out that the presence of a companion causes all metal lines to appear weaker due to line dilution. This means that it leads to an apparent sub-solar metallicity content if a single-star analysis is used. While speculative, this could explain the sub-solar metallicity derived through our GSSP fit.

To test the possible presence of a stripped He star, we calculated a synthetic spectrum with the non-LTE Potsdam Wolf-Rayet (PoWR) model atmosphere code (Hamann & Gräfener 2003, Gräfener et al. 2002) adopting typical parameters from Goetberg et al. (2018) and Vink et al. (2017) for a helium star with $M$ = 4 M$_\odot$ (log $L$ = 4 [L$_\odot$], $T^*$ = 65,000 K, log $Mdot$ = -7.3 M$_\odot$ yr$^{-1}$, hydrogen mass fraction = 0.3). For the B-type star, we used the same TLUSTY model as above. With these parameters, the stripped star is relatively bright in the visual. Due to its temperature, it is seen most easily in He II lines such as the He II 5412 line. As Fig. A6 illustrates, the presence a stripped star should have been easily identified in the spectrum.

Additionally, any star initially massive enough to produce a sufficiently massive stripped He star would have a total lifetime significantly shorter than that of the B-type companion and should already have undergone core-collapse. These arguments makes a stripped-star companion a very unlikely explanation for the properties of the LB-1 system. Therefore, we can relatively safely conclude that the hidden companion is not a stripped He star.

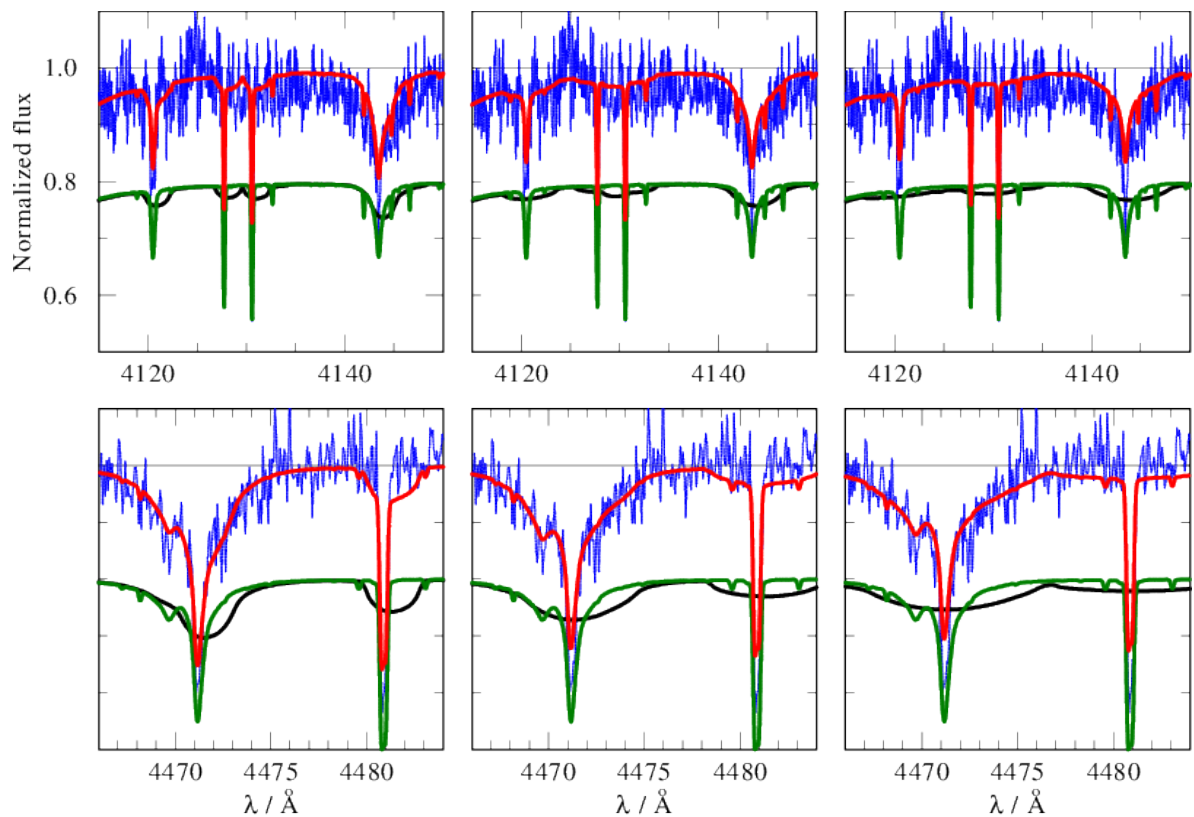

**Fig. A5:** Comparison between normalised HERMES spectrum (blue) and composite synthetic spectrum (red), comprising two TLUSTY models with identical parameters and light contribution (see text) corresponding to the visible B-star (green) and a hypothetical rapidly rotating star of a comparable mass (black) with $v_{eq} \sin i$ = 100 , 200, and 300 km s$^{-1}$ (left, middle, and right panels, respectively). Individual TLUSTY models (green and black) are shifted by -0.2 for clarity.

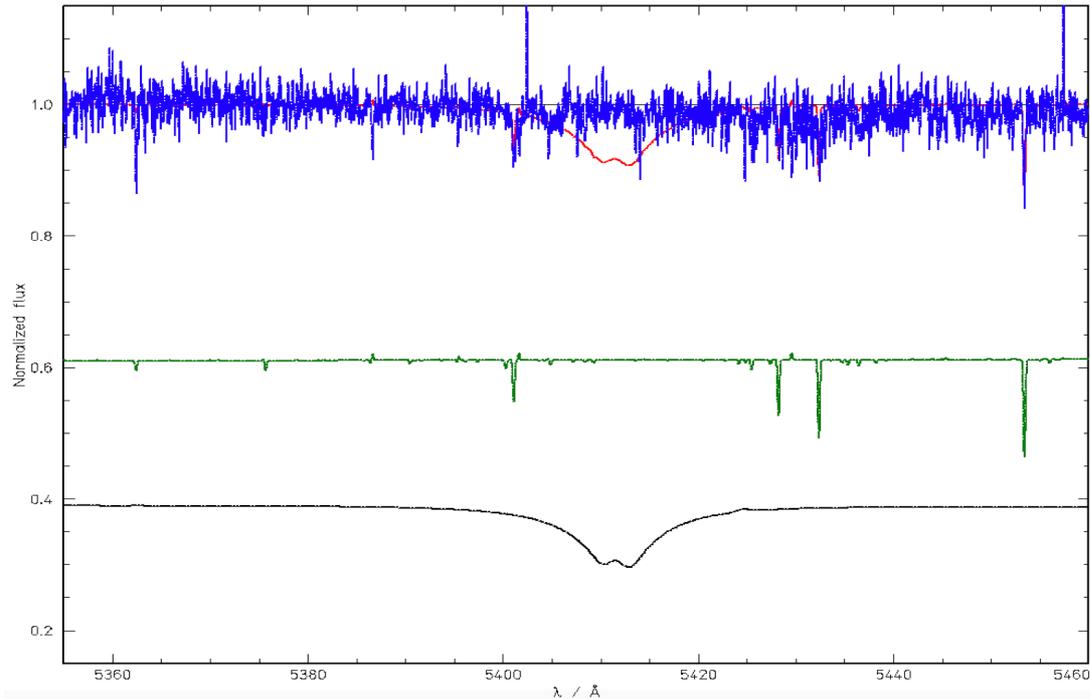

**Fig. A6:** Comparison between normalised HERMES spectrum (blue) and composite synthetic spectrum (red), comprising of our TLUSTY B-star model and a PoWR model calculated for a 4 solar mass stripped star. The Figure illustrates that the hidden companion is unlikely to be a He stripped star.

## Acknowledgements

We acknowledge support from the FWO-Odysseus program under project G0F8H6N. This project has received funding from the European Research Council under European Union's Horizon 2020 research and innovation programme (grant agreement numbers 772225: MULTIPLES and 670519: MAMSIE).

## Author contributions

This paper is based on an original idea by M.A-M. J.B. prepared the observations. LV was the observer. H.V.W. and G.R. performed the HERMES data reduction. J.B. and L.M. performed the atmospheric analysis. M.A-M., K.D., G.B. and T.S. studied the impact of the mixed profile on the RV measurements. D.M.B., M.F., L.M., T.S., A.T. and H.S. contributed to various aspects of the data analysis. All authors contributed to the discussion and interpretation of the results and commented on the written draft of the paper.